\documentstyle[aps,prb]{revtex}
\preprint{}
\begin{document}
\newcommand{\simgeq}{\raisebox{-0.6 ex}{$\stackrel{>}{\sim}$}}
\title{Super-reflection of light from a random amplifying medium with disorder in
the complex refractive index : Statistics of fluctuations}
\author{S. Anantha Ramakrishna$^\dagger$, E. Krishna Das$^\ddagger$, 
G.V. Vijayagovindan$^\ddagger$ and N.Kumar$^\dagger$}
\address {${^\dagger}$Raman Research Institute, C.V. Raman Avenue, Bangalore - 560 080, India. \\
${^\ddagger}$ School of Pure and Applied Physics, Mahatma Gandhi University, 
Priyadarshini Hills (P.O.), Kottayam - 686 560, India.}
\date{\today}
\maketitle
\begin{abstract}
The probability  distribution of the reflection coefficient for light reflected 
from a one-dimensional random amplifying medium with {\it cross-correlated}
spatial disorder in the real  and the imaginary parts
of the refractive index is derived using the method of invariant 
imbedding. The statistics of fluctuations have been obtained 
for both the correlated telegraph noise and the Gaussian white-noise
models  for the disorder. In both cases, an enhanced backscattering
(super-reflection with reflection coefficient greater than unity)
results because of coherent feedback due to Anderson localization
and coherent amplification in the medium. The results show that the effect of  
randomness in the imaginary part of the refractive index on localization 
and super-reflection is qualitatively different.   \\ 
\end{abstract}

Light propagation in disordered media and the associated Anderson localization 
of a wave in both the active as well as the passive random media 
\cite{sheng,pradhan,zhang,paaschens}has been studied extensively. 
In recent years, there has been 
increased interest in light propagation and lasing in active 
random media supported by the several 
experiments carried out on these systems \cite{lawandy,wiersma1,wiersma2,siddique,prasad}.
However the experimental findings of a narrowed spectral emission \cite{lawandy,prasad}
and a pulse narrowing of the emission \cite{wiersma1,siddique} above a well defined threshold
of pumping could be explained merely as an effect of the long diffusive pathlengths in a random
medium with gain and the consequent amplified spontaneous emission (ASE)
\cite{wiersma1,john}. More recently, 
the observed super-narrowing of the emitted spectra from strongly scattering semiconducting 
powder \cite{cao} and from weak scatterers dispersed in high gain organic media \cite{frolov} 
has been attributed to coherent feedback caused by recurrent multiple scattering 
\cite{wiersma3}. It is, however, still debatable if the wave amplification is due to the 
predicted synergy between wave confinement by Anderson localization and coherent 
amplification \cite{pradhan}. \\

In all these studies, the active random medium is considered to scatter the 
propagating wave (light) due to fluctuations in the real part of the refractive index
$(\eta_{r})$ (real potentials) while the coherent amplification is modelled by a phenomenological
spatially constant imaginary part of the refractive index $(\eta_i)$. However, it 
would be of interest to examine the effect of a spatially fluctuating imaginary part of 
the refractive index as well. More so, as the
scattering micro-particles  (e.g. polystyrene microspheres, $TiO_2$ rutile particles) used
in the experiments are not active, a corresponding mismatch 
in the imaginary part of the refractive index is
found to exist. It has been pointed out by Rubio and Kumar \cite{rubio} that a mismatch 
in the imaginary part of the refractive index (imaginary potential) would always 
cause a concomitant reflection (scattering) in addition to the absorption or amplification. 
Mismatch in  $\eta_i$ alone  in an amplifying medium (negative imaginary potential) 
with no mismatch in $\eta_r$ can cause resonant enhancement of the scattering coefficients. 
In fact, the reflection and the transmission coefficients can even diverge as can be seen 
from the simple example of a single imaginary $\delta$  potential in one dimension. 
This would correspond to the experimental situation where the scatterers  
(polystyrene microspheres, say) are suspended in a fluid with the same $\eta_r$ 
(index matching fluid) in which a laser dye is dissolved and optically pumped. 
The scattering caused by the fluctuations in $\eta_i$ would, therefore be expected to 
have non-trivial effects on the wave propagation in the medium.  \\

The transmittance across a randomly amplifying and absorbing
chain was recently considered by Sen \cite{sen} numerically and was shown
to decay exponentially with the increase in length of 
the chain, presumably due to localization. But the 
effects of the fluctuation in the imaginary 
part of the refractive index on lasing in such random media has not been studied so far.
In this work, we consider the statistics of the non-self-averaging fluctuations of 
the reflection coefficient for light incident on a one-dimensional active random medium
with spatial correlated disorder in both the imaginary part as well as the 
real part of the refractive index. \\

We consider a one-dimensional active disordered medium of 
length $L$ with a random complex refractive 
index $\eta, 0 \le x \le L$. 
For simplicity, polarization effects are neglected and light is assumed to be a scalar 
wave. A physical realization of interest here would be an $Er^{3+}$ doped
and pumped, polarization maintaining optical fibre intentionally disordered along its length.
Further, only the linear case of the gain/absorption being independent of the wave
amplitude is considered and non-linear features such as gain saturation are not considered.
Here we would like to emphasize that our treatment is for the possibility of 
super-reflection ($r >1$) {\it i.e.,} for an amplifier and not an 
oscillator\cite{soukolis1,expln}.
The complex wave amplitude $E(x)$ obeys the Helmhotz equation 
inside the medium,
\begin{equation}
\label{wave}
\frac{d^{2} E(x)}{d x^{2} } + k^{2} \left( 1 + \eta(x) \right) E(x) = 0
\end{equation}
where k is the wave vector in the medium ($k^{2} = \omega^{2}/c^{2} \epsilon_{0}$) and $\eta(x)
= \eta_{r}(x) + \imath [\bar{\eta_{i}} + \eta_{i}(x)]$ is the complex refractive index. 
Here $\eta_{r}(x)$ and $\eta_{i}(x)$ are random and $\bar{\eta_{i}}$ is a constant representing 
the average amplification or absorption in the medium according as 
$\bar{\eta_{i}}$ is negative or positive. It is well known that 
Eq.(\ref{wave}) can be transformed to give an equation for the  evolution of 
the emergent quantity, namely,   
the complex amplitude reflection coefficient $R(L) = [r(L)]^{1/2} \exp[\imath \theta (L)]$ 
as a function
of the sample length $L$, via the method of invariant imbedding \cite{kumar,rammal} as 
\begin{equation}
\label{reflect}
\frac{dR(L)}{dL} = 2 i k R(L) + \frac{ik}{2} \eta(L) \left[ 1+ R(L) \right]^{2}
\end{equation}
Equation.(\ref{reflect}) is a stochastic differential equation and we are interested 
in the corresponding Fokker-Planck equation for the probability distribution 
$P(r,\theta;L)$ which can be readily obtained following the standard procedures. Thus,
let $\Pi(r,\theta;L)$ be the density of points in the $(r,\theta)$ phase space. 
Now  $\Pi(r,\theta;L)$ must satisfy 
the Stochastic Liouville equation \cite{vankampen}, and by the van Kampen 
lemma\cite{vankampen}, the 
probability distribution function $ P(r,\theta;L) = \langle \Pi (r,\theta; L) 
\rangle_{\eta_{r},\eta_{i}}$, where the angular brackets denote averaging over 
all the realizations of the random refractive indices
$\eta_{r}$ and $\eta_{i}$. 

\subsection{The Gaussian $\delta$ correlated (white-noise) disorder}

First, let us consider the simplest case namely that of a Gaussian $\delta$-correlated
(white-noise) model.
In this model, $\eta_{r}$ and $\eta_{i}$ are assumed to have  $\delta$ 
correlated Gaussian distributions with $\langle \eta_{r}(L)\rangle = 0$, 
$\langle \eta_{i}(L)\rangle = 0$,
$\langle \eta_{r}(L) \eta_{r}(L') \rangle =  \Delta_{r}^{2} \delta(L-L')$ and   
$\langle \eta_{i}(L) \eta_{i}(L') \rangle =  \Delta_{i}^{2} \delta(L-L')$. 
This model would most appropriately describe the 
case of a continuous random medium such as a laser-dye doped gel or intralipid 
suspension \cite{tanosaki}, where the fluctuations in $\eta_{r}$ and $\eta_{i}$ 
are uncorrelated. Using the Novikov theorem \cite{novikov} to average over all 
configurations of $\eta_{r}$ and $\eta_{i}$, we obtain in the 
Random Phase Approximation (RPA) ({\it i.e.,} $ P(r,\theta) = P(r)/2 \pi$), 
\begin{equation}
\label{wneqn}
\frac{\partial P}{\partial l} = \phi_{r} {\bf L_{R}}P  + \phi_{i}{\bf L_{I}}P 
 + 2 A \frac{\partial (rP)} {\partial r} ,    
\end{equation}
 where the linear operators $ {\bf L_{R}}$ and ${\bf L_{I}}$ are given by
\begin{eqnarray}
\label{rpaop1}
{\bf L_{R}} & = & \frac{1}{2} \left[ r (r-1)^{2} \frac{\partial^{2}}
{\partial r^{2}} + (5 r^{2} - 6 r +1)
\frac{\partial}{\partial r} + 2 (2r-1) \right] , \\
\label{rpaop2}
{\bf L_{I}} & = & \frac{1}{2} \left[ r(r^{2}+10r+1) \frac{\partial^{2}}
{\partial r^{2}} + (5 r^{2}+30r+1)
\frac{\partial}{\partial r} +2(2r+5) \right] ,
\end{eqnarray}
and the non-dimensional sample length $l = 1/2 max\{\Delta_{r}^{2},
\Delta_{i}^{2}\} k^{2} L \equiv L/l_{c}$,
$ \phi_r = \Delta_{r}^{2}/ max\{\Delta_{r}^{2},\Delta_{i}^{2}\}$, $ \phi_r 
= \Delta_{i}^{2}/ max\{ \Delta_{r}^{2},\Delta_{i}^{2} \}$ and 
$ A = 2 \bar{\eta_{i}}/ max \{\Delta_{r}^{2},\Delta_{i}^{2}\}k \equiv l_{c}/l_{amp}$ 
Here $l_{amp} = (\bar{\eta_{i}} k)^{-1}$ is the amplification length in the 
medium defined by the average 
of the imaginary part of the refractive index and $max$ implies the superior value 
of the arguments. The RPA is known to be valid in the  
the weak disorder limit, $k l_{c} >> 1$, where $l_{c}$ is the localization length
\cite{rammal}. We point out that even if $\eta_{r}$ and $\eta_{i}$ were 
cross-correlated, the final equations do not differ in the RPA for the white-noise
model (because $\langle {\bf L_{1} L_{2}} P\rangle_{\theta} = 0$ see equations (9),(10)). 
	
The asymptotic $ l \rightarrow \infty$ limiting solution of Equation.(\ref{wneqn}) 
obtained by setting $\partial P / \partial l = 0$ is given by,
\begin{eqnarray}
\label{wnsoln}
P(r;\infty) & = & P_{0} \frac{\exp \left\{ -2A/\gamma \tan^{-1}\left[\left( 
(\phi_{r}+ \phi_{i})r +5\phi_{i}-\phi_{r} \right)/\gamma \right] \right\} }
{\left[  (\phi_{r}+ \phi_{i}) (1+r^{2})+ 2(5\phi_{i}-\phi_{r})r 
\right] }~~~~~~~~~~~~~~~~~~~~~~~~~~\phi_{r}>2\phi_{i}    \\
& = & \frac{P_{0}}{(\phi_{r}+ \phi_{i}) (1+r^{2})+ 2(5\phi_{i}-\phi_{r})r}
\left[ \frac{ (\phi_{r}+ \phi_{i})r +5\phi_{i}-\phi_{r} -\gamma}{ (\phi_{r}+ \phi_{i})r 
+5\phi_{i}-\phi_{r} +\gamma} \right]^{-A/\gamma} 
~~\phi_{r}   <   2\phi_{i} \nonumber
\end{eqnarray}
where $\gamma = \sqrt{12 \phi_{i} \vert \phi_{r}-2\phi_{i}\vert }$ 
and $P_{0}$ is a normalization 
constant given by 
$[\int_{0}^{\infty} P(r,\infty) dr]^{-1}$. The limit $l \rightarrow \infty$ 
implies physically  $ L >> l_{c}$. This expression goes over straightforwardly
to the result of Pradhan and Kumar\cite{pradhan} in the  limiting 
case of pure real disorder ($\phi_{i} = 0$).
Thus the statistics qualitatively differ in the two regimes for an 
amplifying medium : (i) when the real part disorder dominates 
($ \phi_{r} > 2 \phi_{i}$) and (ii) when the
Imaginary part disorder dominates ($ \phi_{r} < 2 \phi_{I}$). \\ 

We have also solved equation(\ref{wneqn}) numerically for finite 
length to investigate the approach to the 
asymptotic forms given by eq.(\ref{wnsoln}). In Fig.(1), the plots of 
$P(r,l)$ for the case of real disorder
dominating ($\phi_{r}> 2 \phi_{i}$) for different lengths of the medium are shown. 
The probability distribution for the case of a pure imaginary mismatch ($\phi_{r} 
= 0$), with the real part $\eta_{r}$ being index-matched is shown in Fig.(2). The 
line joining the dots in both the figures corresponds to the asymptotic $P(r;\infty)$ 
solution. In the case of amplifying medium, the value of reflectivity 
($r_{max}$) at which $P(r;l)$ peaks increases with the average value of the 
amplification factor $\vert A \vert$.  
For the case of imaginary part disorder  dominating, $P(r;l)$ has a peak 
at small values of the reflectivity even for moderate values of the 
amplification. In the case of an absorbing medium with the
imaginary diorder dominating, the probability 
distribution has a monotonic decreasing behaviour and is maximum at $r=0$.
A finite probability of reflection at $r>1$ in the
absorbing case and at $r<1$ in the
amplifying case ($A<0$) is recognised to be a consequence of 
the two-sidedness of the white-noise process for the complex 
refractive index, which allows the imaginary part of the refractive index
($\bar{\eta_{i}} + \eta_{i}$) to take on locally both positive and
negative values for any given value of the average. It should be 
noted that this limiting form of $P(r,\infty)$ gives a
weak logarithmic divergence for $\langle r\rangle$ (for $\phi_{i}
\ne 0$ ), regardless of the sign of $A$ for
both absorption and amplification. Thus amplification has a much more drastic
effect on the reflectivity than attenuation. The white-noise process allows the 
local fluctuations in $\eta_i$ to be very large and the effect of a finite 
mean value $\bar{\eta_{i}}$ is small. It is thus a case of the
fluctuations dominating over the mean.
We also find that the numerical solutions 
saturate to the limiting forms for $l$\simgeq$1$. 
So most of the reflection occurs 
from within a localization length. This enhanced backscattering is quite different 
from that caused by light diffusion\cite{wiersma1,john}.  In 
the latter case, the distribution of optical path length, because of 
exponential growth of wave amplitude due to coherent amplification 
in one-dimension, gives $P^{D} (r;\infty) \sim \ln (r)^{1/2}/r$ for $r>>1$. 
This decays much slower than the $P(r;\infty)$ for $r\rightarrow 
\infty$, as given by eq.(\ref{wnsoln}).  \\

\subsection{Correlated telegraph disorder} 
In the case of the white-noise disorder, the imaginary part of the refractive 
index was allowed
to take on both positive and negative values {\it i.e.,} the medium could 
be locally both amplifying or absorbing. With a view to studying purely 
amplifying/absorbing random media, we use the telegraph disorder model to 
describe the fluctutations in the refractive index. Moreover, since 
the gain/absorption coefficient is physically always bounded from above, 
the fluctuations in the imaginary part of the refractive index 
are better described by this dichotomic Markov process 
({\it i.e.,} spatial Telegraph noise). 
Further, we recognize that in discrete random media such as 
microparticles suspended in a laser dye solution used in experiments, 
the real and the imaginary parts of the refractive index fluctuate 
spatially in the same manner and can, therefore, be described by the same 
stochastic process. A telegraph noise with a finite correlation 
length is most appropriate to describe such a situation. Accordingly, 
we will choose $\eta_{r}(L) = 
\alpha \chi (L)$ and $\eta_{i}(L) = \beta \chi(L)$ with an  average value for 
the imaginary part $\bar{\eta_{i}}$.  Here $\chi(L)$ is taken to 
be a dichotomic markov process which can take on the values $\pm\chi$ 
such that $\langle \chi(L) \rangle = 0$ and $\langle \chi(L) \chi(L') 
\rangle = \chi^{2} \exp\left(-\Gamma \vert L-L' \vert \right)$,where
$\Gamma^{-1}$ is the correlation length in the medium. \\

Now, defining as before, $P(r,\theta;L) = \langle \Pi(r,\theta;L) 
\rangle_{\chi}$ and $W(r,\theta;L) = \langle \chi(L) \Pi(r,\theta;L) 
\rangle_{\chi}$, and using the ``formulae of differentiation'' of Shapiro 
and Loginov\cite{shapiro} to average over the dichotomous configurations of 
$\chi(L)$, we obtain  
\begin{eqnarray}
\frac{\partial P}{\partial L} & = & -2k\frac{\partial P}{\partial \theta} +
\bar{\eta_{i}} {\bf L_{2}} P + (\alpha {\bf L_{1}} + \beta {\bf L_{2}}) W , \\
\frac{\partial W}{\partial L} & = & \chi^{2}(\alpha {\bf L_{1}} + \beta {\bf L_{2}}) 
P - 2k\frac{\partial W}{\partial \theta} + \bar{\eta_{i}} {\bf L_{2}} W 
- \Gamma W,
\end{eqnarray}
where the linear operators ${\bf L_{1}}$ and ${\bf L_{2}}$ are :
\begin{eqnarray}
{\bf L_{1}} = -k\left[ \sin\theta \frac{\partial}{\partial r} \sqrt{r}
(1-r) + \frac{\partial}{\partial \theta} + \frac{1}{2} \left(\sqrt{r} +\frac{1}
{\sqrt{r}}\right) \frac{\partial}{\partial \theta} \cos\theta \right] \\
{\bf L_{2}} = k \left[ \cos\theta \frac{\partial}{\partial r} \sqrt{r}
(1+r) + 2 \frac{\partial}{\partial r} r+\frac{1}{2} \left(\sqrt{r} -\frac{1} 
{\sqrt{r}}\right) \frac{\partial}{\partial \theta} \sin\theta \right]
\end{eqnarray}
We thus get a closed system of equations for  $P(r,\theta,L)$ and  
$W(r,\theta,L)$. These equations 
go over correctly to the corresponding eq.(\ref{wneqn}) 
in the white-noise limit obtained by taking the limit $\chi^{2} 
\rightarrow \infty$, $\Gamma \rightarrow \infty$ while keeping 
$\chi^{2} /\Gamma = \Delta^{2}$ constant. In this limit,
the equation for $P(r,\theta;L)$ becomes autonomous {\it i.e.,} it 
gets decoupled from $W(r,\theta;L)$. \\
 
In the RPA and in the asymptotic
limit $L \rightarrow \infty$, these equations simplify to 
\begin{eqnarray}
\label{tneqn1}
\beta \bar{\eta_{i}} {\bf L_{I}} P + \alpha^{2} {\bf L_{R}} W + 
\beta^{2} {\bf L_{I}} W  = 0  ~~,  \\
\label{tneqn2}
\alpha^{2} {\bf L_{R}} P + \beta^{2} 
{\bf L_{I}} P + 2 A \frac{\partial (rP)}{\partial r} - 
\frac{\bar{\eta_{i}} \beta}{\chi^{2}} {\bf L_{R}} W = 0
\end{eqnarray}
where ${\bf L_{R}}$ and ${\bf L_{I}}$ are given by eq.(\ref{rpaop1}) and 
eq.(\ref{rpaop2}) and $ A = 2 \Gamma \bar{\eta_{i}}/\chi^{2}$.  
Interestingly in the case of the pure real part disorder 
($\beta =  0 $), the form of the 
telegraph noise equation for $P(r;\infty)$ is identical to that for the white-
noise case, but with the coefficient $A = 2\Gamma \bar{\eta_{i}}/k\chi^{2}$. 
Similiarly,  in the case of the pure imaginary part disorder 
($\alpha =  0 $), the form of the
telegraph noise equation for $P(r;\infty)$ is again identical to 
that for the white-noise case, but with the coefficient
$A = 2\Gamma \bar{\eta_{i}}/k(\chi^{2}-\bar{\eta_{i}})$.
However, for $\beta \chi < \vert \bar{\eta_{i}} \vert$, the imaginary part 
of the refractive index is always positive (absorbing) or negative (amplifying).
Hence the solution for these two cases is also given by eq.(\ref{wnsoln}), 
the solutions being 
valid in the interval $ 0 < r< 1$ for the absorbing medium, and 
$1<r<\infty$ for the amplifying medium. Outside the intervals, the probability 
density $P(r;L)$ vanishes. \\

A complete solution for the eq.(\ref{tneqn1})and eq.(\ref{tneqn2})
is obtained as   
\begin{eqnarray}
\label{tgexp1}
P(r;\infty) & = & P_{0}\left[ \frac{1}{\xi_{+}(1+\zeta_{+} r+r^{2})} +
\frac{1}{\xi_{-}(1+\zeta_{-} r+r^{2})} \right] \exp [-2A 
\left(I_{+}(r)+I_{-}(r)\right)] \\
I_{\pm}(r) & = & \frac{1}{\xi_{\pm} \sqrt{\zeta_{\pm}^{2}-4}} \ln 
\left\vert \frac{r - r_{\pm}^{(2)}}{r-r_{\pm}^{(1)}}
\right\vert ~~~~~~~~~~~~~~\vert \zeta_{\pm}\vert > 2 \nonumber\\
 & = & \frac{1}{\xi_{\pm} \sqrt{\zeta_{\pm}^{2}-4}} \tan^{-1} 
\left(\frac{\zeta_{\pm}+2r}
{\sqrt{\zeta_{\pm}^{2}-4}} \right) ~~~~~~~~~~~~~~~~~ \vert \zeta_{\pm} \vert < 2 
\nonumber
\end{eqnarray}
where $\xi_{\pm} = \alpha^{2} +\beta^{2}\pm\beta\bar{\eta_{i}}/\chi$, 
$ \zeta_{\pm} =[10(\beta^{2}\pm\beta\bar{\eta_{i}}/\chi)
-2\alpha^{2}]/[1\pm\sqrt{\beta}+\alpha^{2}] $, $r_{\pm}^{(1)} = -1/2[\zeta_{\pm}
+(\zeta_{\pm}^{2} -4)^{1/2}]$, $r_{\pm}^{(2)} = -1/2[
\zeta_{\pm}-(\zeta_{\pm}^{2} -4)^{1/2}]$ 
and $P_{0}$ is a normalization coefficient. This expressions become 
the same as given by eqn.(\ref{wnsoln}) in the white noise limit
($\chi^{2} \rightarrow 0$, $\Gamma \rightarrow \infty$ and $\chi^{2}/\Gamma$
being constant).\\ 

The solutions for one-sided disorder in the imaginary part exhibit
three qualitatively different behaviours corresponding to choices of the 
parameters $\alpha$, $\beta$
and $\bar{\eta_{i}}$ ($\chi$ is an arbitrary constant and can be set to unity 
without loss of generality). First, we note that the case of 
$\alpha^{2}+\beta^{2}-\beta\bar{\eta_{i}}/\chi = 0$, corresponds to a singular 
perturbation of the differential equation for $P(r;\infty)$. This condition 
can be interpreted as a threshold condition by noting that the localization 
length is given by 
$l_{c}^{-1} \sim (\alpha^{2} +\beta^{2})$ and the effective amplification 
length is given by $l_{amp}^{-1} = \beta \bar{\eta_{i}}$. This condition 
then corresponds to a matching of length scales in the problem, $l_{c} = l_{amp}$.
In the regime where the amplification dominates the localization 
($\alpha^{2}+\beta^{2}-\beta\bar{\eta_{i}} < 0$ or $l_{c} > l_{amp}$), 
the solutions exhibit  a monotonic decreasing behaviour in 
the region of interest ($1 \leq r < \infty$). Here the disorder in the real part 
($\alpha$) is small and does not affect the statistics appreciably, as can be seen 
from Fig.(3a). For ($\alpha^{2}+\beta^{2}-\beta\bar{\eta_{i}} > 0$ or 
$l_{c} < l_{amp}$), a natural boundary arises for the solutions of the equation
at $r_{-}^{(2)}$ which falls in the domain of physical interest 
($1 \leq r < \infty$). Now the solutions given by the expression(\ref{tgexp1})
are valid in the range $r_{-}^{(2)} \leq r < \infty$ with $P(r;\infty) = 0$ 
outside.  In this regime the localization dominates ($l_{c} < l_{amp}$),
if $2A/[\xi_{-}(\zeta_{-}^{2}-4)^{-1/2}] >1$ and we have a broad distribution with 
peak at $r_{max} > r_{-}^{(2)}$ and $P(r_{-}^{(2)};\infty) = 0$ (Fig.(3b)). 
The value of $r_{max}$ is large for small disorder in the real part  
($\alpha^{2}+\beta^{2}-\beta\bar{\eta_{i}}~
$\simgeq$~0$), and decreases as 
$\alpha$ increases. The behaviour in this region is dominated by the disorder
in the real part of the refractive index. A third qualitatively different 
behaviour occurs for $l_{c} < l_{amp}$ and $2A/[\xi_{-}(\zeta_{-}^{2}-4)^{-1/2}] <1$. 
Then the expression given by eqn.(\ref{tgexp1}) diverges at $r_{-}^{(2)}$.
This divergence is, however, normalizable implying that $P(r;\infty)$ 
is peaked (in fact, sharply) at that point.
This behaviour can be readily understood by noting that  
the second condition which can be 
rewritten as $\bar{\eta_{i}}^{2} (\Gamma / k)^2 < 3\beta (\vert
\bar{\eta_{i}} \vert -\beta ) [\alpha^{2}+2\beta (\vert\bar{\eta_{i}} 
\vert -\beta )] $, is basically a condition on the correlation length
($l_{corr} = \Gamma^{-1}$). This condition is satisfied for small $\Gamma
$ (large $l_{corr}$). Then the reflection is essentially from a single
potential barrier and thus has a sharply defined value. It should be noted 
that, as $\alpha \rightarrow \infty$, $P(r;\infty) \rightarrow
\delta (r-1)$, as expected. \\

The solutions for the case of a two-sided disorder for the imaginary part ($\beta > 
\vert \bar{\eta_{i}} \vert$) are similiar to the 
solutions for the white noise case. It should be noted that there
does not exist real $r_{-}^{(2)}$ which falls into the physical region of interest
($0 \leq r < \infty$). In this case the large disorder in the imaginary part 
($\beta$) causes the effects of localization to dominate. 
However, in all cases of amplification, 
for a finite A and $\alpha^{2}+\beta^{2}-\beta\bar{\eta_{i}} \neq 0$, there is 
a universal $1/r^{2}$ tail for the $P(r;\infty)$.
For the case of pure imaginary disorder($\alpha = 0 $), we similiarly see a 
monotonically decreasing behaviour of $P(r;\infty)$ with $r$ for
one-sided disorder ($\beta < \vert\bar{\eta_{i}}\vert~$ or $~l_{c} > l_{amp}$) (Fig.(3c)), 
and a $P(r;\infty)$ with a peak for 
two-sided disorder ($\beta > \vert \bar{\eta_{i}}\vert~$ or $~l_{c} < l_{amp}$) (Fig.(3d)).
With increase in $\beta$ for two-sided disorder, the epak shifts to 
smaller values of reflectivity as the effects of absorption show up, 
until for large enough $\beta$, the peaks occurs at $r=0$ and we 
again have a monotonically decreasing $P(r;\infty)$.
It should be mentioned that all these effects are seen 
for the case of absorption also, with the roles of $r_{-}^{(1)}$ and 
$r_{-}^{(2)}$ interchanged. \\ 

In conclusion, we have studied the statistics of super-reflection from a one-dimensional
disordered system with spatial randomness both in the real and the imaginary parts of 
the complex refractive index. We have discussed the models of disorder qualitatively  
applicable to experimental systems such as intentionally disordered optical fibres
with gain ($Er^{3+}$-doped) and obtained the probability distribution function 
of the reflectivity for the cases of a white-noise disorder and a 
correlated telegraph disorder.
In both cases, an enhanced reflection results because of coherent feedback 
due to Anderson localization and coherent amplification.
In the case of white-noise disorder, the statistics are qualitatively different in the
two regimes of the real part disorder dominating ($\Delta_{r}^{2} > 2\Delta_{i}^{2}$) 
and the imaginary part disorder dominating ($\Delta_{r}^{2} < 2\Delta_{i}^{2}$).
In the case of telegraph disorder, we obtain three qualitatively different behaviours 
for $P(r;\infty)$ depending on  threshold conditions involving the localization 
length, the amplification length and the correlation length.
Thus the fluctuation in the imaginary 
part of the refractive index is seen to have a non-trivial and qualitatively 
different  effect on localization and lasing from such random media. 
Finally, as the phenomenon considered here is concerned with the issue of statistical 
fluctuations (noise) in a random amplifying medium, we propose for it the acronym
RAMAN (Random Amplifying Medium And Noise). \\  

\section*{acknowledgements}
One of us (N.K) thanks Douglas Stone for suggesting the acronym RAMAN

\vspace{5mm}
\begin{center} {\bf FIGURE CAPTIONS} \\ \end{center}

\noindent Fig. 1. The probability distribution of reflectivity $P(r;l)$ in the case of 
the white noise disorder given by eqn.(\ref{wneqn}) and the real disorder dominating 
($\phi_{r} = 1.0$, $\phi_{i} = 0.1$),
for the different sample lengths indicated. The line joining the dots is the 
analtic result for $P(r;\infty)$. The amplification parameter is $A = -0.25$ .\\

\noindent Fig. 2. The probability distribution $P(r;l)$ in the case of the white noise
disorder given by eqn.(\ref{wneqn}) and a pure imaginary mismatch ($\phi_{r} = 0$) for 
different lengths of the sample. The line joining the dots is the 
analtic result for $P(r;\infty)$. The amplification parameter is $A = -1$. \\

\noindent Fig. 3. The probability distribution $P(r;l)$ in the case of the correlated
telegraph noise. (a) $l_{c} > l_{amp}$ and (b) $l_{c} < l_{amp} $ 
are for one-sided disorder ($\beta < \vert \bar{\eta_{i}} \vert$)
with disorder in both the real and the imaginary parts.
(c) $l_{c} > l_{amp}$ and (d) $l_{c} < l_{amp}$ are for two-sided 
disorder ($\beta > \vert \bar{\eta_{i}} \vert$) and pure imaginary
mismatch ($\alpha = 0$).


\begin{references}
\bibitem{expln}In the 1-D case, there are no extended states irrespective of the 
sign of $\bar{\eta_{i}}$ and hence, the probability of infinite gain (oscillations)
is always zero even though the average of the gain can be infinite because of the long 
tail of the distribution of reflectivity $P(r)$. The energy storage in the disordered system
with a intensity-independent gain can, however, become infinite. Indeed, the non self
averaging nature of the reflection makes the problem different from the case of a 
non-random system, where the infinite gain can be obtained deterministically for
certain resonant conditions of the sample length {\it etc.} In contrast for the 
random amplifying medium, the threshold condition  will demand that the porbability for 
infinite gain be non-zero which is not satisfied in the 1-D case treated here.
\end{references}
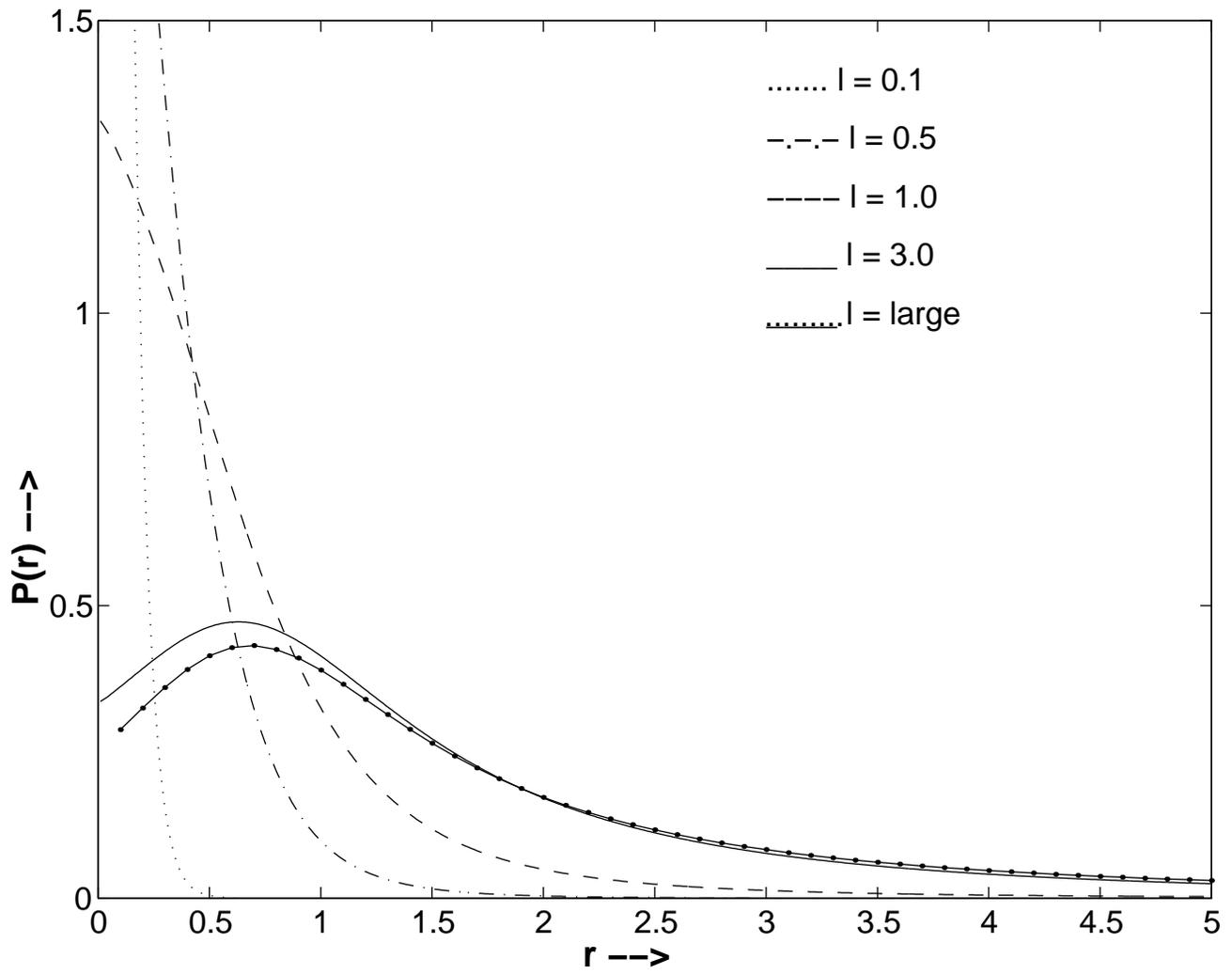
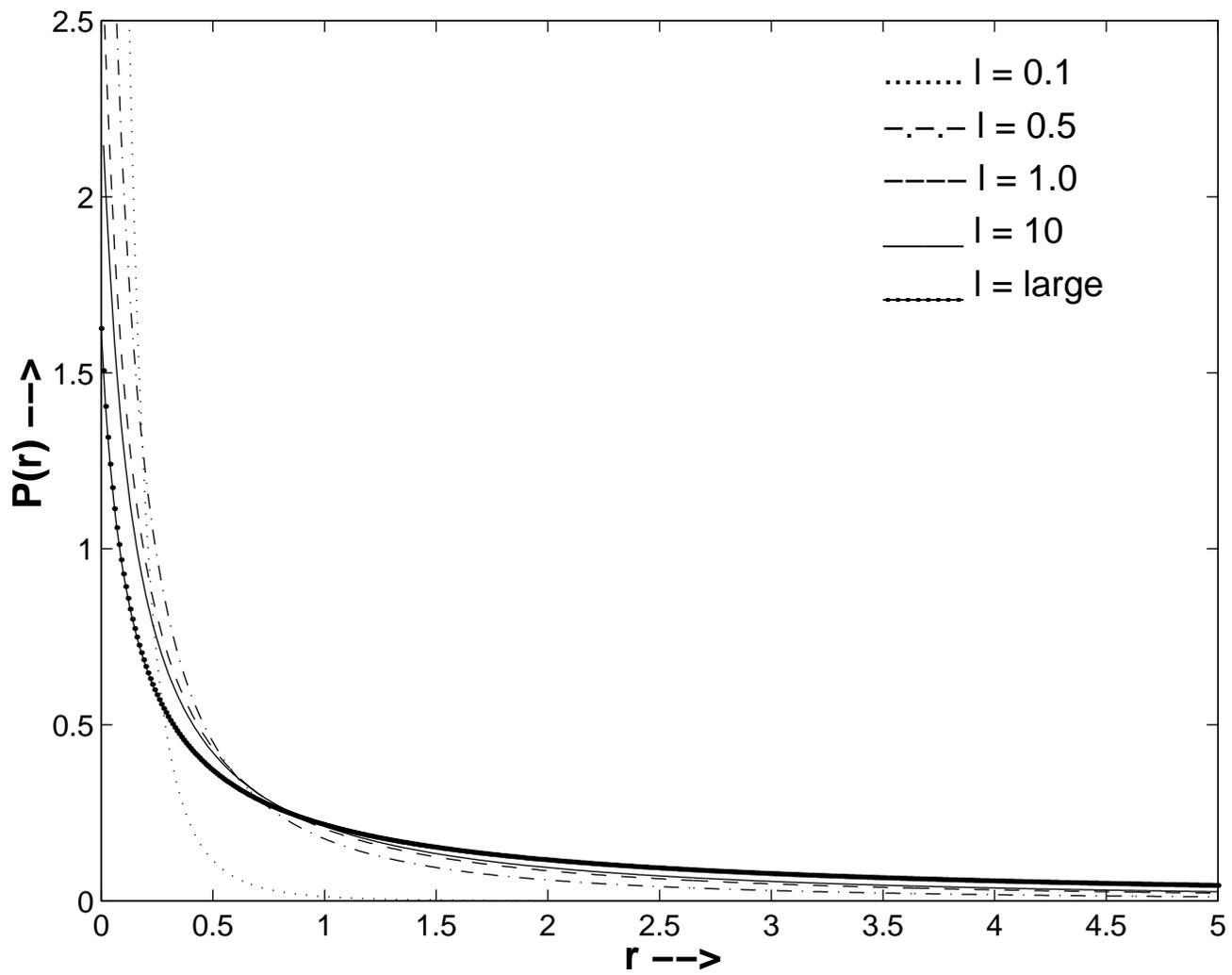
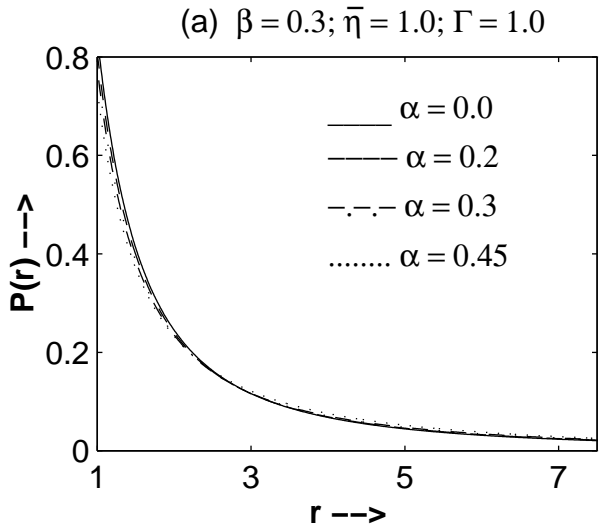
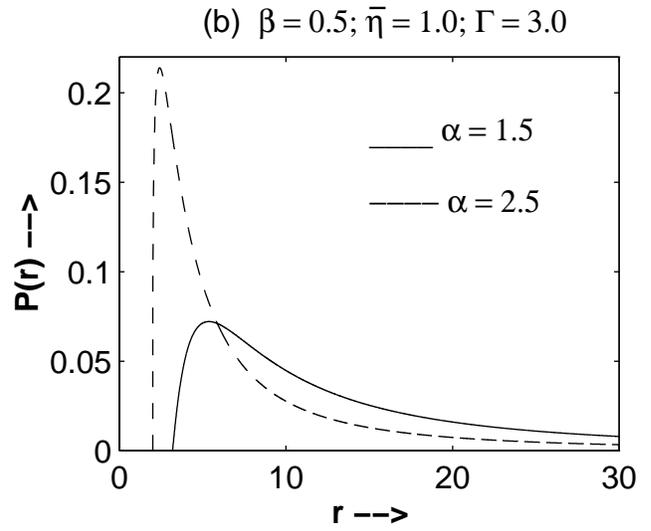
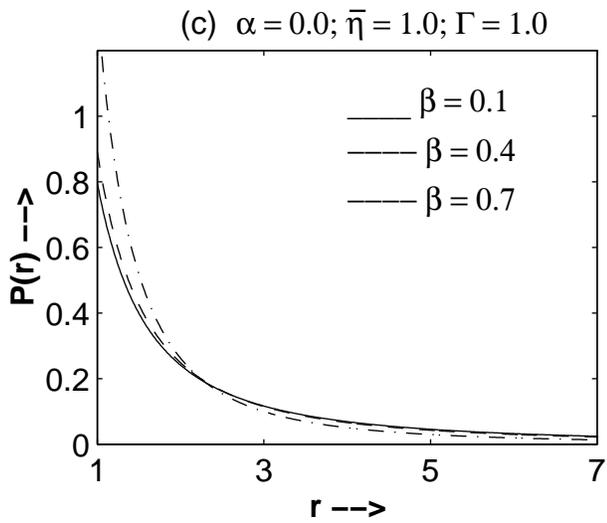
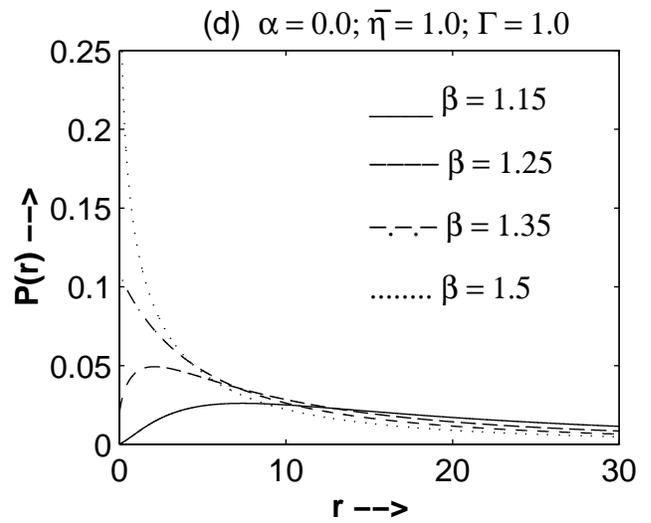
\end{document}